\pdfoutput=1

\documentclass[showpacs,aps,prl,twocolumn,superscriptaddress]{revtex4}
\usepackage{graphicx} 
\usepackage{dcolumn}
\usepackage{bm}
\usepackage{amssymb,amsmath}
\usepackage{epstopdf}
\usepackage[T1]{fontenc}
\usepackage[latin9]{inputenc}
\usepackage{color}
\usepackage{float}
\usepackage{graphicx}
\usepackage{esint}
\usepackage{bm}
\usepackage{graphics}
\makeatletter
\@ifundefined{textcolor}{}
{%
 \definecolor{BLACK}{gray}{0}
 \definecolor{WHITE}{gray}{1}
 \definecolor{RED}{rgb}{1,0,0}
 \definecolor{GREEN}{rgb}{0,1,0}
 \definecolor{BLUE}{rgb}{0,0,1}
 \definecolor{CYAN}{cmyk}{1,0,0,0}
 \definecolor{MAGENTA}{cmyk}{0,1,0,0}
 \definecolor{YELLOW}{cmyk}{0,0,1,0}
}

\begin{document}
\title{Superradiant Decay of Cyclotron Resonance of Two-Dimensional Electron Gases}
\normalsize

\author{Qi Zhang}
\affiliation{Department of Electrical and Computer Engineering, Department of Physics and Astronomy, and Department of Materials Science and NanoEngineering, Rice University, Houston, Texas 77005, USA}

\author{Takashi Arikawa}
\thanks{Present address: Department of Physics, Kyoto University, Japan}
\affiliation{Department of Electrical and Computer Engineering, Department of Physics and Astronomy, and Department of Materials Science and NanoEngineering, Rice University, Houston, Texas 77005, USA}

\author{Eiji Kato}
\affiliation{Advantest America, Inc., Princeton, New Jersey, 08540, USA}

\author{John L.~Reno}
\affiliation{Sandia National Laboratories, CINT, Albuquerque, New Mexico 87185, USA}
\author{Wei Pan}
\affiliation{Sandia National Laboratories, Albuquerque, New Mexico 87185, USA}

\author{John D.~Watson}
\author{Michael J.~Manfra}
\affiliation{Department of Physics, School of Materials Engineering, School of Electrical and Computer Engineering, and Birck Nanotechnology Center, Purdue University, West Lafayette, Indiana 47907, USA}

\author{Michael A.~Zudov}
\affiliation{School of Physics and Astronomy, University of Minnesota, Minneapolis, Minnesota 55455, USA}

\author{Michail Tokman}
\affiliation{Institute of Applied Physics, Russian Academy of Sciences, 603950 Nizhny Novgorod, Russia}

\author{Maria Erukhimova}
\affiliation{Institute of Applied Physics, Russian Academy of Sciences, 603950 Nizhny Novgorod, Russia}

\author{Alexey Belyanin}
\affiliation{Department of Physics and Astronomy, Texas A\&M University, College Station, Texas 77843, USA}

\author{Junichiro Kono}
\thanks{Author to whom correspondence should be addressed}
\email[]{kono@rice.edu}
\affiliation{Department of Electrical and Computer Engineering, Department of Physics and Astronomy, and Department of Materials Science and NanoEngineering, Rice University, Houston, Texas 77005, USA}

\date{\today}

\begin{abstract}
We report on the observation of collective radiative decay, or superradiance, of cyclotron resonance (CR) in high-mobility two-dimensional electron gases in GaAs quantum wells using time-domain terahertz magnetospectroscopy.  The decay rate of coherent CR oscillations increases linearly with the electron density in a wide range, which is a hallmark of superradiant damping.  Our fully quantum mechanical theory provides a universal formula for the decay rate, which reproduces our experimental data without any adjustable parameter.  These results firmly establish the many-body nature of CR decoherence in this system, despite the fact that the CR frequency is immune to electron-electron interactions due to Kohn's theorem.
\end{abstract}


\maketitle

Understanding and controlling the dynamics of superposition states is of fundamental importance in diverse fields of quantum science and technology~\cite{Wineland13RMP,AwschalometAl13Science,ArndtHornberger14NP}.  In particular, how an excited many-body system relaxes remains one of the fundamental questions in nonequilibrium statistical mechanics~\cite{Zwanzig01Book,KinoshitaetAl06Nature,PolkovnikovetAl11RMP}.  A Landau-quantized, high-mobility two-dimensional electron gas (2DEG) provides a uniquely clean and tunable solid-state system in which to explore coherent many-electron dynamics.  A superposition of massively degenerate Landau levels (LLs) can be created by a coherent terahertz (THz) pulse through cyclotron resonance (CR) absorption~\cite{HiltonetAl12Book}. How rapidly the coherence of this many-body superposition state decays has not been well understood.  Even though the CR frequency, $\omega_\mathrm{c}$, is immune to many-body interactions due to Kohn's theorem~\cite{Kohn61PR}, the decoherence of CR can be affected by electron-electron interactions.  

Theoretical studies predicted that the linewidth of CR should oscillate with the LL filling factor since the screening capability (i.e., the density of states at the Fermi energy) of a 2DEG oscillates with the filling factor~\cite{Ando75JPSJ,Ando77JPSJ,DasSarma81PRB,LassnigGornik83SSC,AndoMurayama85JPSJ,FoglerShklovskii98PRL}.  However, despite several decades of experimental studies of CR in 2DEGs using continuous-wave and incoherent methods~\cite{EnglertetAl83SSC,SchlesingeretAl84PRB,HeitmannetAl86PRB,EnsslinetAl87PRB,SeidenbuschGornik87PRB,BatkeetAl88PRB,RichteretAl89PRB,KonoetAl94PRB}, no clear evidence for the predicted CR linewidth oscillations has been obtained for high-mobility, high-density samples, partly due to the `saturation effect'; i.e., in the high-conductivity limit, the 2DEG behaves as a metallic mirror, reflecting most of the incident light at the CR peak, resulting in an undesirable broadening of transmittance linewidths~\cite{ChiuetAl76SS,ChouetAl88PRB,Mikhailov04PRB}.

\begin{figure}
\includegraphics[scale=0.4]{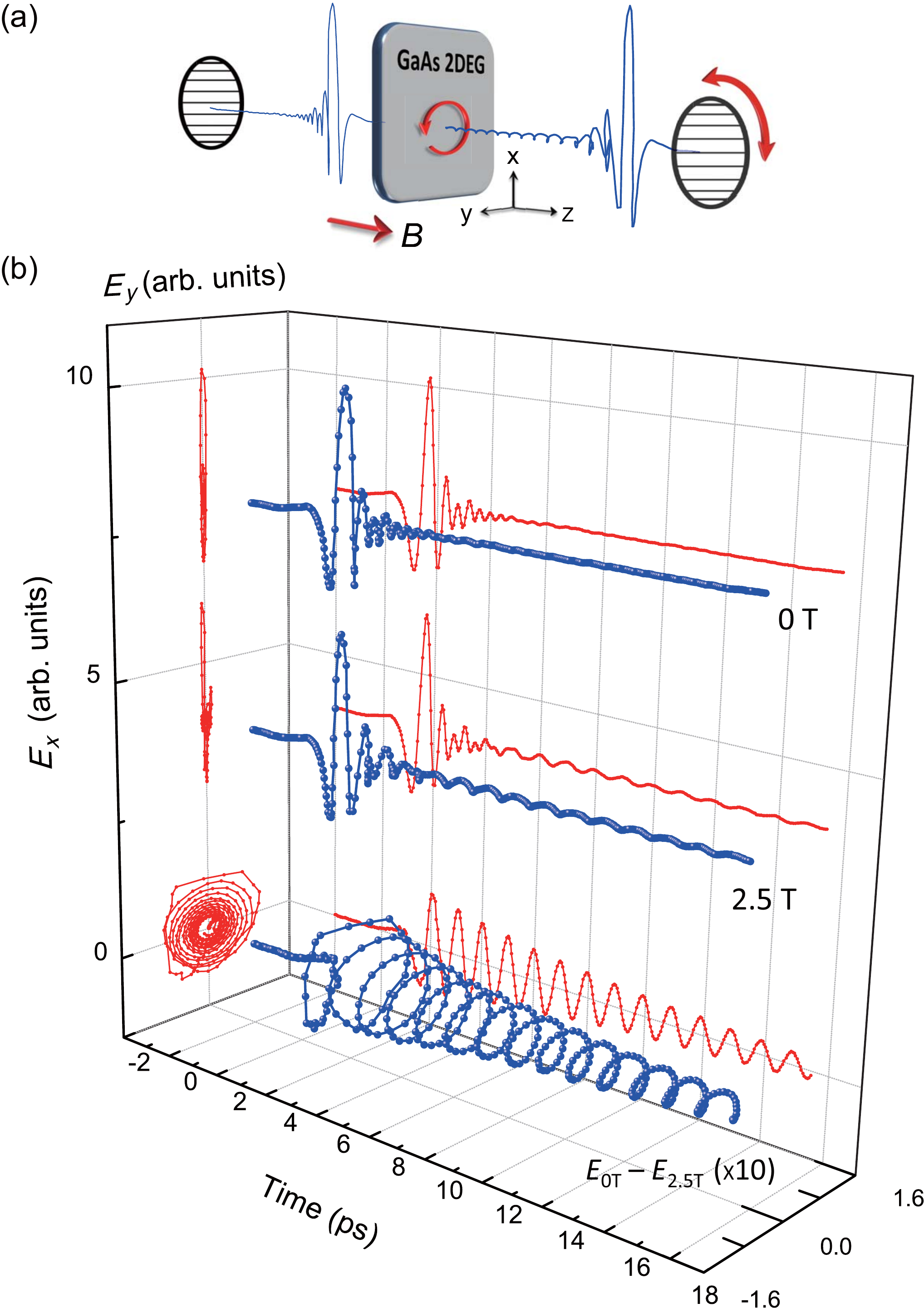}
\caption{(Color online) (a)~A schematic of the polarization-resolved THz magnetotransmission experiment in the Faraday geometry.  (b)~Coherent cyclotron resonance oscillations in the time domain.  Each blue dot represents the tip of the THz electric field at a given time.  The red traces are the projections of the waveforms onto the $E_x$-$t$ and $E_x$-$E_y$ planes.  The bottom trace is the difference between the top (0~T) and middle (2.5~T) traces.}
\label{fig1}
\end{figure}

Here, we performed a systematic study on CR decoherence in high-mobility 2DEGs by using time-domain THz magnetospectroscopy~\cite{WangetAl07OL,ArikawaetAl11PRB}, measuring the CR decay time, $\tau_\mathrm{CR}$, as a function of temperature ($T$), magnetic field ($B$), electron density ($n_e$), and mobility ($\mu_e$). As $T$ decreases, $\tau_\mathrm{CR}$ increases due to reduced electron-phonon interaction, but $\tau_\mathrm{CR}$ eventually saturates at low $T$. The low-$T$ saturation value of $\tau_\mathrm{CR}$ is uncorrelated with $\mu_e$; rather, the CR decay rate $\Gamma_\mathrm{CR}$ ($\equiv$ $\tau_\mathrm{CR}^{-1}$) increases linearly with $n_e$.  We developed a fully quantum mechanical theory for describing coherent CR, which clearly identifies superradiant (SR) damping~\cite{Dicke54PR,Haken84Book} to be the dominant decay mechanism.  Namely, $\Gamma_\mathrm{CR}$ is dominated by \emph{cooperative radiative decay} at low $T$, which is much faster than any other phase-breaking scattering processes.

We studied two samples of modulation-doped GaAs quantum wells grown by molecular beam epitaxy.  Sample 1 had $n_e$ and $\mu_e$ of 1.9 $\times$ 10$^{11}$~cm$^{-2}$ and 2.2 $\times$ 10$^6$~cm$^2$/Vs, respectively, in the dark, while after illumination at 4~K they changed to 3.1 $\times$ 10$^{11}$~cm$^{-2}$ and 3.9 $\times$ 10$^6$~cm$^2$/Vs;  intermediate $n_e$ values were achieved by careful control of illumination times.  Sample 2 had $n_e$ = 5 $\times$ 10$^{10}$~cm$^{-2}$ and $\mu_e$ = 4.4 $\times$ 10$^6$~cm$^2$/Vs. 

Time-domain THz magnetospectroscopy experiments were performed using two different systems.  One system used a Ti:Sapphire regenerative amplifier (Clark MXR, Inc.)~with 775~nm center wavelength, $1$~kHz repetition rate, and 150~fs pulse width to generate and detect THz pulses with ZnTe crystals~\cite{WangetAl07OL,ArikawaetAl11PRB}.  The other system  (TAS7500TS, Advantest Corp.) utilized two ultrashort fiber lasers with the electronically controlled optical sampling technique to generate and detect THz waveforms with 132~ps scan range and 8~ms single scan time; attached compact fiber-coupled photoconductive switch emitter and detector allowed us to couple the THz beam into the magnet with minimum effort.  The incident beam was linearly polarized by the first polarizer, and by rotating the second polarizer, the transmitted THz field was measured in both $x$- and $y$-directions [Fig.~\ref{fig1}(a)].  Figure \ref{fig1}(b) shows transmitted THz waveforms in the time domain.  Each blue dot represents the tip of the THz electric field, $\mathbf{E}$ = ($E_x$,$E_y$), at a given time.  The red traces are the projections of the waveforms onto the $E_x$-$t$ plane and $E_x$-$E_y$ plane.  The top and middle traces show the transmitted THz waveforms at 0~T and 2.5~T, respectively. The 2.5~T trace contains long-lived oscillations with circular polarization.  The bottom trace is the difference between the two, $E_\mathrm{0 T}(t)-E_\mathrm{2.5 T}(t)$, which is proportional to the THz-induced current at the CR frequency of the 2DEG [see Eq.~(\ref{bound})].  Hence, its decay time $\tau_\mathrm{CR}$ can be directly and accurately determined through fitting with  $A\exp(-t/\tau_\mathrm{CR})\cdot\sin(\omega_\mathrm{c}t+\phi_0)$, where $A$ and $\phi_0$ are the CR amplitude and the initial phase, respectively.

\begin{figure}
\includegraphics[scale=0.57]{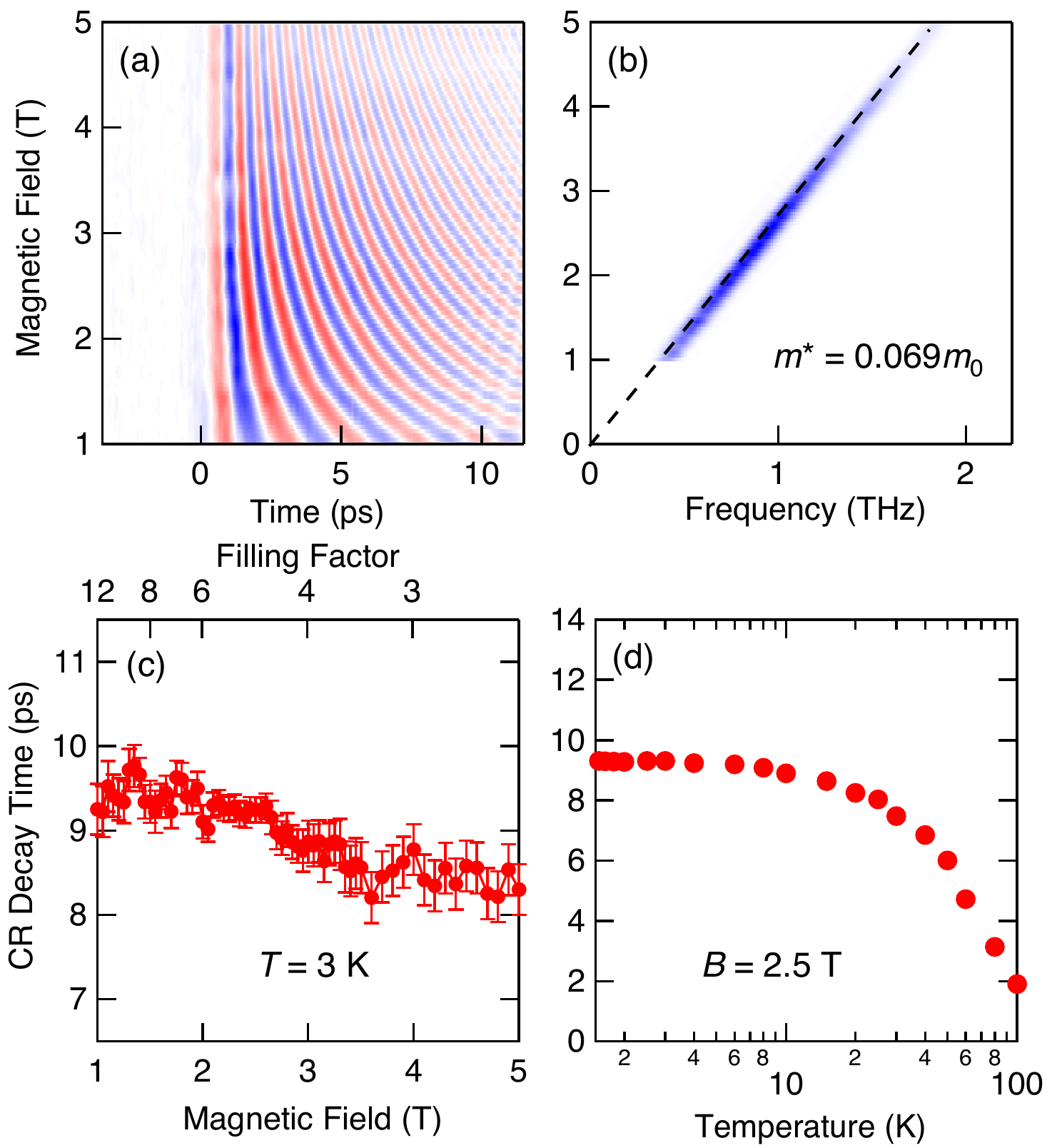}
\caption{(Color online) (a)~Magnetic field dependence of CR oscillations, showing peaks (blue) and valleys (red). (b)~The frequency-domain version of (a). Black dashed line: linear fit with a cyclotron mass of 0.069$m_0$.  (c)~Magnetic field dependence of $\tau_\mathrm{CR}$ at 3~K. (d)~Temperature dependence of $\tau_\mathrm{CR}$ at 2.5~T.  All the data are for Sample 1.}
\label{fig2}
\end{figure}

Figure~\ref{fig2}(a) shows CR oscillations at various $B$ for Sample 1 after illumination.  The inter-LL spacing, or $\hbar\omega_\mathrm{c}$, increases with $B$. Figure \ref{fig2}(b) shows the Fourier transform of the time-domain data in Fig.~\ref{fig2}(a) into the frequency domain.  A linear $B$ dependence of $\omega_\mathrm{c}$ = $eB/m^*c$ provides electron cyclotron mass $m^*$ = 0.069$m_0$, where $m_0$ = 9.11 $\times$ 10$^{-28}$~g.  As shown in Fig.~\ref{fig2}(c), the variance of $\tau_\mathrm{CR}$ with $B$ is small; $\tau_\mathrm{CR}$ slightly decreases with increasing $B$, but no oscillatory behavior is observed.  Figure \ref{fig2}(d) shows that $\tau_\mathrm{CR}$ increases with decreasing $T$ but saturates at $\sim$9.5~ps when $T \lesssim$ 10~K.

\begin{figure}
\includegraphics[scale=0.42]{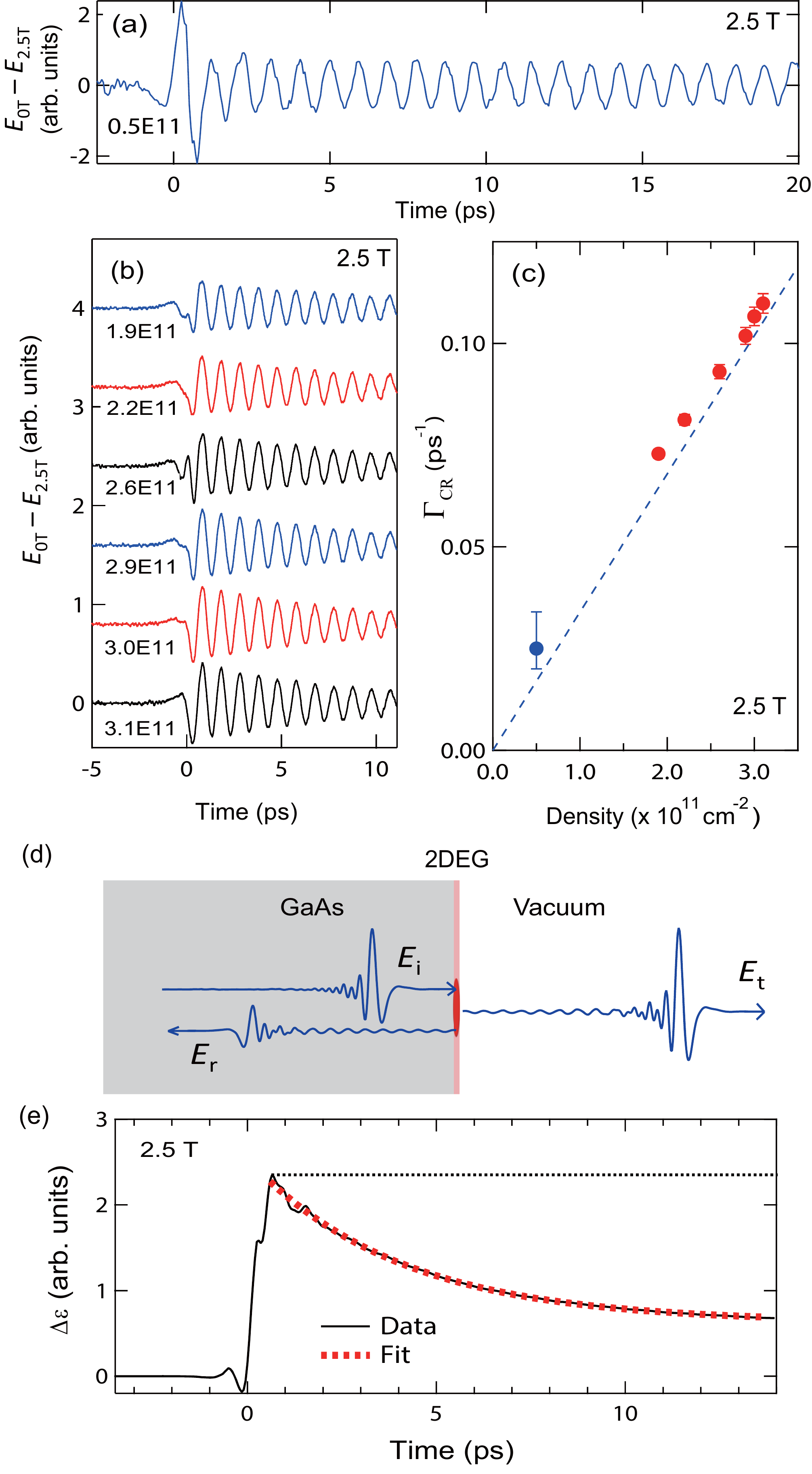}
\caption{(Color online) (a)~Low-density sample (Sample 2) exhibiting the longest $\tau_\mathrm{CR}$ of $\sim$40~ps. (b)~CR oscillations in Sample 1 with different densities by controlling the illumination time. (c)~Decay rate as a function of density. Blue solid circle: Sample 2.  Red solid circles: Sample 1.  The blue dashed line represents Eq.~(\ref{Gamma-SR}) with no adjustable parameter. (d)~Incident $E_\mathrm{i}$, reflected $E_\mathrm{r}$ and transmitted $E_\mathrm{t}$ THz pulses at the 2DEG. (e)~The decay of the THz-pulse-excited energy in the 2DEG. The red dashed line is an exponential fit. About 80\% of the energy relaxes through CR superradiance.}
\label{fig3}
\end{figure}

The values of $\tau_\mathrm{CR}$ at low $T$ were much shorter than the DC scattering time, $\tau_{\rm DC}$ = $m^*\mu_e/e$, of the same samples.  Furthermore, there was no correlation between $\tau_\mathrm{CR}$ and $\tau_{\rm DC}$; in some cases, higher-mobility samples revealed shorter $\tau_\mathrm{CR}$ values. 
On the other hand, $\tau_\mathrm{CR}$ showed strong correlation with $n_e$.  As $n_e$ was increased, $\tau_\mathrm{CR}$ was found to decrease in a clear and reproducible manner, as shown in Figs.~\ref{fig3}(a) and \ref{fig3}(b). The low-density sample (Sample 2) exhibited the longest $\tau_{\rm CR}$ value of $\sim$40~ps.  Figure \ref{fig3}(c) shows that the decay rate, $\Gamma_\mathrm{CR}$, increases linearly with $n_e$, which, as described below, is consistent with SR damping of CR.

A qualitative picture is as follows.  A coherent incident THz pulse induces a polarization in the 2DEG, i.e., macroscopic coherence as a result of \emph{individual cyclotron dipoles oscillating in phase}.  The resulting free induction decay of polarization occurs in a SR manner, much faster than the dephasing of single oscillators.  The SR decay rate, $\Gamma_\mathrm{SR}$, is roughly $N$ times higher than the individual radiative decay rate, where $N \sim n_e \lambda^2$ is the number of electrons within the transverse coherence area of the incident THz wave with wavelength $\lambda$.  In an ultraclean 2DEG, $\Gamma_\mathrm{SR}$ is higher than the rates of all other phase-breaking scattering mechanisms.  This scenario explains not only the $n_e$ dependence of  $\tau_\mathrm{CR}$ but also its weak $B$ dependence as well as the saturation of $\tau_\mathrm{CR}$ at low $T$.

Furthermore, the SR nature of CR emission not only dramatically speeds up the radiative decay but also makes CR radiation more directional and collinear with the excitation pulse.  Thus, most of the CR radiation could be collected, allowing us to analyze the incident and radiated THz waves quantitatively.  At the 2DEG, shown in Fig.~\ref{fig3}(d), the incident ($E_\mathrm{i}$), reflected ($E_\mathrm{r}$), and transmitted ($E_\mathrm{t}$) THz fields satisfy the boundary condition, $E_\mathrm{i}(t) + E_\mathrm{r}(t) = E_\mathrm{t}(t)$.  With the full knowledge of $E_\mathrm{t}(t)$ at 0~T and 2.5~T as well as $\sigma_\mathrm{0 T}(\omega)$, the optical conductivity of the 2DEG at 0~T, we obtained both $E_\mathrm{i}(t)$ and $E_\mathrm{r}(t)$ at 2.5~T. The THz-induced energy increase in the 2DEG, $\Delta\varepsilon(t)$, is proportional to $\int_0^t(n_{\rm GaAs}\left|E_{\rm i}(t')\right|^2-n_{\rm GaAs}\left|E_{\rm r}(t')\right|^2-\left|E_{\rm t}(t')\right|^2)dt'$, shown in Fig.~\ref{fig3}(e). 
If the energy is dissipated nonradiatively, i.e., via scattering, $\Delta\varepsilon(t)$ would be a step function, as indicated by the black dotted line in Fig.~\ref{fig3}(e). However, our data instead show that most of the absorbed energy goes back into the field, again supporting the SR picture.  By fitting $\Delta\varepsilon(t)$ with an exponential with a baseline, we found that the majority ($\sim$80\%) of the energy decays radiatively; the other 20\% could be due to imperfect collection and any residual scattering loss. 

We developed a quantum mechanical model for THz excitation and coherent CR emission of a 2DEG in a perpendicular $B$, valid for an excitation pulse of an arbitrary duration with respect to $\Gamma_\mathrm{CR}^{-1}$ and $\omega_\mathrm{c}^{-1}$.
We start from the master equation for the density operator in the coordinate representation, $d\hat{\rho}/dt = - (i/\hbar) [ \hat{H},\hat{\rho} ] + \hat{R}(\hat{\rho})$,
%
%
where $\hat{R}(\hat{\rho})$ is the relaxation operator. The Hamiltonian for an electron of mass $m^*$ in a confining potential $U(\textit{\textbf{r}})$ interacting with an optical and magnetic field described by the vector potential $\mathbf{A} = \mathbf{A}_\mathrm{opt} + \mathbf{A}_{B}$ is
 \begin{equation}
 \label{ham}
\hat{H} = \frac{\hat{p}^2}{2m^*} +U({\bf r}) - \frac{e}{2m^*c^2} \left(\mathbf{A} \mathbf{\hat{p}} + \mathbf{\hat{p}} \mathbf{A} \right) + \frac{e}{2m^*c^2} \mathbf{A}^2, 
\end{equation}
where $\mathbf{\hat{p}} = - i \hbar \nabla$. In our case, the energy of the first-excited quantum-well subband is much higher than all energy scales relevant to the problem, and so we can assume that the electrons stay in the ground subband.  Furthermore, in our case of relatively modest $B$ and low-energy excitations, we can neglect any band nonparabolicity, and thus, the resulting LLs are equally spaced.

Care should be exercised in choosing the correct form of $\hat{R}(\hat{\rho})$.  A standard empirical expression for the relaxation of the off-diagonal elements of the density matrix, $R_{mn} = \gamma_{mn} \rho_{mn}$, can be used only within the rotating wave approximation (RWA) and under the assumption that the relaxation rate, $\gamma$, is much smaller than eigen-frequencies of the system. Neither of these approximations is valid in our case of an ultrashort excitation pulse and frequencies in the (sub)THz range. As was shown in Refs.~\onlinecite{TokmanErukhimova13JL} and \onlinecite{Tokman09PRA}, outside the RWA the standard relaxation term leads to spurious terms in the equations for quantum-mechanical averages of the dipole moment and populations, including violation of a standard relationship, $\mathbf{j} = \dot{\mathbf{d}}$,  between quantum mechanical averages of the current density ($\mathbf{j}$) and dipole moment ($\mathbf{d}$). 

Following Ref.~\onlinecite{TokmanErukhimova13JL}, we choose the relaxation operator in the coordinate representation and for $\mathbf{A}$ = 0 as  
\begin{equation}
\label{relax}
\hat{R} = -\gamma_{\perp} \left( \hat{\rho}_{\perp} - \hat{\rho}_{\perp}^T \right) - \gamma_{\parallel} \left( \hat{\rho}_{\parallel} - \hat{\rho}_{\parallel,0} \right),
\end{equation}
where $\hat{\rho}_{\perp,\parallel}$ are the off-diagonal and diagonal components of the density operator, respectively, with corresponding transverse ($\gamma_{\perp}$) and longitudinal ($\gamma_{\parallel}$) relaxation rates, and $\hat{\rho}_{\parallel,0}$ is an equilibrium distribution of populations.  For $\mathbf{A} \neq 0$, Eq.~(\ref{relax}) has to be transformed to preserve gauge invariance as specified in Ref.~\onlinecite{TokmanErukhimova13JL}. Using the master equation and Eq.~(\ref{ham}), we can derive a set of equations for the quantum mechanical averages of $\mathbf{j}$, $\mathbf{d}$, and energy density ($W$) of the system: 
\begin{align}
\label{main}
\ddot{\mathbf{d}} + 2 \gamma_{\perp} \dot{\mathbf{d}} + \frac{e}{m} \overline{\nabla U \hat{\rho}} + \omega_c \mathbf{b}\times \dot{\mathbf{d}}  &= \frac{e^2}{m}\, \overline{\hat{\rho}}\,\mathbf{E}(t), \\ 
\mathbf{j} &= \dot{\mathbf{d}}, \\
\dot{W} + \gamma_{\parallel} (W - W_0) &= \dot{\mathbf{d}} \mathbf{E}(t),
\end{align}
where  $\mathbf{d}$ = $-e \overline{\mathbf{r}\hat{\rho}}$, $\mathbf{j}$ = $-(e/m) \overline{(\hat{\mathbf{p}} + e \mathbf{A}/c)\hat{\rho}}$, $W$ = $\overline{ \left( \hat{\mathbf{p}}^2/2m + U(\mathbf{r}) \right) \hat{\rho}}$, $W_0$ is an equilibrium energy density, $\mathbf{b}$ is a unit vector along $\mathbf{B} \parallel \mathbf{z}$, and $\mathbf{E}$ is the THz electric field. The overbar means taking the trace with the density matrix,  i.e., $\overline{\hat{\mathbf{g}} \hat{\rho}} = \int \hat{\mathbf{g}}(\mathbf{r}) \rho(\mathbf{r}, \mathbf{r}') \delta(\mathbf{r}-\mathbf{r}') \, d^3 r d^3 r'$~\cite{LandauLifshitz76Book}. 

Since all electrons are in the ground subband, i.e., effectively 2D with no confinement potential transverse to $\mathbf{B}$, we can drop the term containing $\nabla U$.  We also assume that the excitation pulse is too weak to perturb populations. Then from Eqs.~(3)-(5) we can obtain the following equation for the circularly polarized current $j_+ = j_x - ij_y$:
\begin{equation}
\label{cur}
{dj_+ \over dt} + (i \omega_c + 2 \gamma_{\perp}) j_+  = \alpha E_+(t), 
\end{equation}
where $E_+$ = $E_x - iE_y$, $\alpha$ = $\omega_p^2/4\pi$, and $\omega_p$ = $\sqrt{4 \pi e^2 \overline{\hat{\rho}}/m^*}$ is the plasma frequency. The electric field acting on the current in Eq.~(\ref{cur}) consists of the 
excitation pulse $\mathbf{E}_0 = (E_{0x}(t), 0, 0)$ and the field radiated by the current, $\mathbf{e} = (e_x, e_y, 0)$. Note that $\overline{\hat{\rho}} = n_e/L$ is the electron volume density, where $L$ is the thickness of the 2DEG layer. There is an interesting characteristic feature of an electron system with a parabolic band, i.e., equally spaced LLs: the current and its radiation are determined by the \emph{total} $n_e$ and independent of how the electrons are distributed among the LLs.  Therefore, our results remain valid even at room temperature and low $B$.

From the boundary conditions on both sides of the 2DEG, i.e., the continuity of the electric field and the jump in the magnetic field $(b_x, b_y, 0)$ radiated by the current, $b_+(z = +0) - b_+(z = -0) = 4 \pi j_+ L/c$, together with Maxwell's equations relating $\mathbf{e}$ and $\mathbf{b}$ fields in the outgoing radiation, we can obtain the radiation field on the surface of the 2DEG, 
\begin{equation}
\label{bound}
e_+ (1 + n_{\rm GaAs}) = -\frac{4 \pi j_+}{c}L,
\end{equation}
where $n_\mathrm{GaAs}$ is the substrate refractive index. This gives the final equation for the current:
\begin{equation}
\label{cur2}
{dj_+ \over dt} + (i \omega_c + \Gamma_\mathrm{CR}) j_+  = \alpha E_{0x}(t), 
\end{equation}
where the CR decay rate, $\Gamma_\mathrm{CR}$, now includes the collective radiative contribution proportional to $n_e$:
\begin{align}
\label{Gamma}
 \Gamma_\mathrm{CR} = \Gamma_\mathrm{scatt.} + \Gamma_\mathrm{SR},
\end{align}
where
\begin{align}
\Gamma_\mathrm{scatt.} &= 2\gamma_\mathrm{\perp}, \\
\Gamma_\mathrm{SR} = \frac{\omega_p^2 L}{(1+n_\mathrm{GaAs})c} &= \frac{4\pi e^2n_e}{m^*(1+n_\mathrm{GaAs})c} \label{Gamma-SR}.
\end{align}
%
%
%
%
%
%
%
As shown by the dashed line in Fig.~\ref{fig3}(c), Eq.~(\ref{Gamma-SR}) reproduces the observed linear $n_e$ dependence of $\Gamma_\mathrm{CR}$ \emph{without any adjustable parameter}, strongly supporting the notion that SR damping dominates the CR decay process in these high-$\mu_e$ samples. 

\begin{figure}
\begin{center}
\includegraphics[scale=0.58]{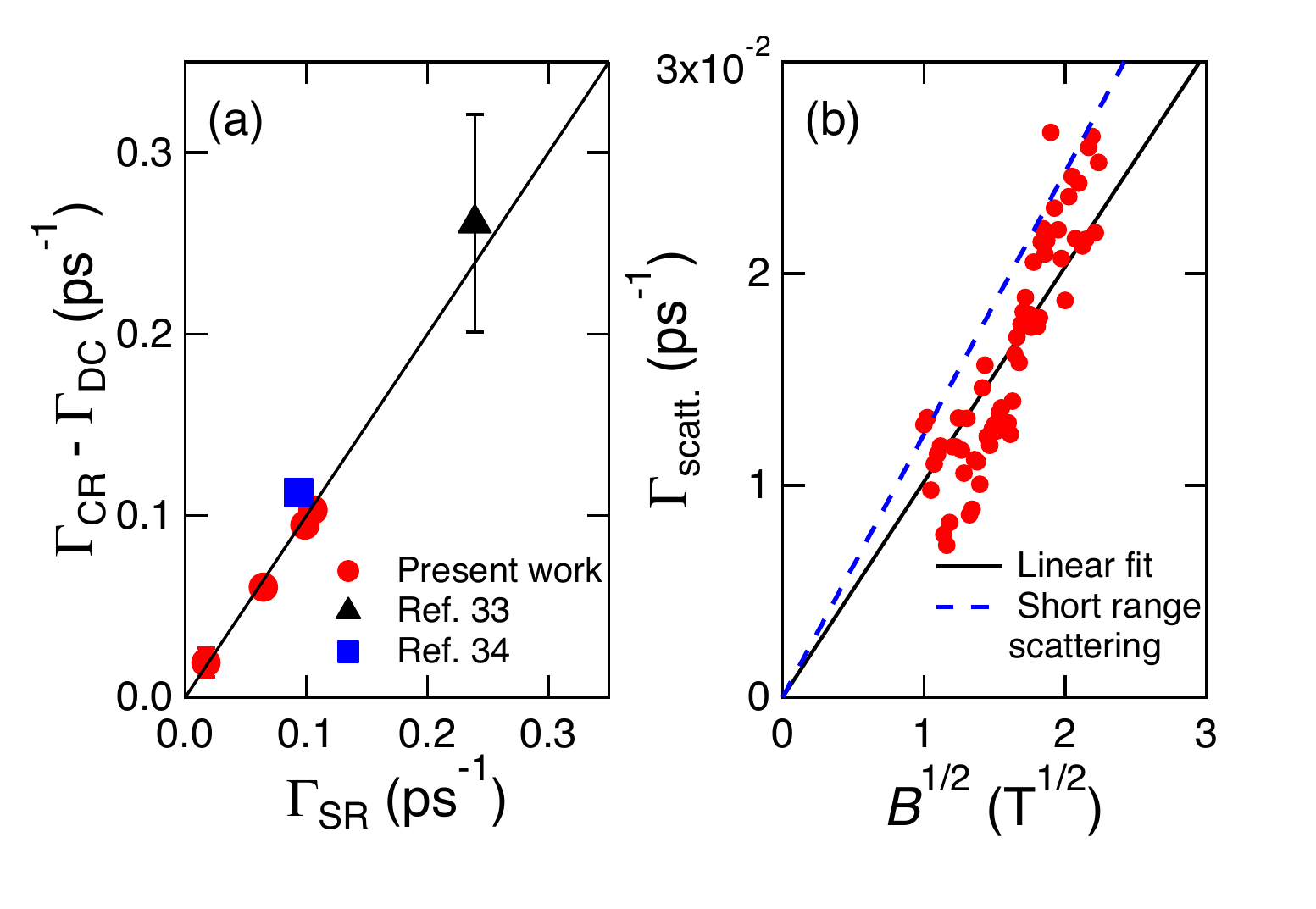}
\caption{(Color online) (a)~The measured values of $\Gamma_\mathrm{CR} - \Gamma_\mathrm{DC}$ versus $\Gamma_\mathrm{SR}$ given by Eq.~(\ref{Gamma-SR}) for four representative data points in the present study and values from Refs.~\cite{SomeNurmikko94APL,IkebeetAl10PRL}. The solid line has a slope of 1.  (b)~$\Gamma_\mathrm{scatt.} \equiv \Gamma_\mathrm{CR} - \Gamma_\mathrm{SR}$ as a function of $\sqrt{B}$.  Red solid line: linear fit.  Blue dashed line: prediction based on short-range scattering~\cite{FoglerShklovskii98PRL}.}
\label{fig4}
\end{center}
\end{figure}

For low-$n_e$ and low-$\mu_e$ samples, $\Gamma_\mathrm{scatt.}$~is not negligible compared to $\Gamma_\mathrm{SR}$, and thus, the values of $\Gamma_\mathrm{CR}$ are expected to deviate from $\Gamma_\mathrm{SR}$, as seen in Fig.~\ref{fig3}(c). For rough estimation, one can assume that $\Gamma_\mathrm{scatt.} \approx \Gamma_\mathrm{DC} = \tau_\mathrm{DC}^{-1} = e/m^*\mu_e$ in Eq.~(\ref{Gamma}). For example, the value of $\tau_\mathrm{CR}$ (= $\Gamma_\mathrm{CR}^{-1}$) estimated in this manner for Sample 2 is $\sim$44~ps, which agrees well with the measured value (40$\pm$10~ps). Figure \ref{fig4}(a) plots $\Gamma_\mathrm{CR} - \Gamma_\mathrm{DC}$ versus $\Gamma_\mathrm{SR}$ for four representative data points in the present study as well as data previously reported for 2DEGs with different values of $n_e$ and $\mu_e$~\cite{SomeNurmikko94APL,IkebeetAl10PRL}.  
The linear relationship with a slope of 1 seen in this plot proves the validity of the following convenient formula 
\begin{align}
\tau_\mathrm{CR} = {m^* \over e} \left [ \frac{4\pi e \,n_e}{(1+n_\mathrm{GaAs})c} + {1 \over \mu_e} \right ]^{-1}
\label{convenient}
\end{align}
for estimating $\tau_\mathrm{CR}$ from the knowledge of $n_e$ and $\mu_e$.

Finally, Eq.~(\ref{Gamma}) allows us to determine $\Gamma_\mathrm{scatt.}$~as $\Gamma_\mathrm{CR} - \Gamma_\mathrm{SR}$.  In particular, we interpret the small but non-negligible $B$-dependence of $\Gamma_\mathrm{CR}$ shown in Fig.~\ref{fig2}(c) to be the $B$-dependence of $\Gamma_\mathrm{scatt}$.  Figure \ref{fig4}(b) shows $\Gamma_\mathrm{CR} - \Gamma_\mathrm{SR}$ versus $\sqrt{B}$, which exhibits an approximately linear relationship, consistent with theoretical predictions based on short-range scattering~\cite{Ando75JPSJ,FoglerShklovskii98PRL}.

In summary, we studied the decay dynamics of Landau-quantized 2DEGs coherently and resonantly excited by THz pulses.  We found that the decay rate of coherent cyclotron oscillations increases linearly with electron density, which we interpret as evidence of superradiance.  Our detailed quantum mechanical calculations confirmed this interpretation, reproducing our experimental observation quantitatively without any adjustable parameter.  Overall, this study demonstrates the cooperative nature of decay dynamics of a quantum-degenerate, interacting electron system, even though its resonant frequency is independent of many-body interactions.

We acknowledge support from the National Science Foundation (Grant Nos.~DMR-1310138 and OISE-0968405).  This work was performed, in part, at the Center for Integrated Nanotechnologies, a U.S. Department of Energy, Office of Basic Energy Sciences user facility.  Sandia National Laboratories is a multi-program laboratory managed and operated by Sandia Corporation, a wholly owned subsidiary of Lockheed Martin Corporation, for the U.S. Department of Energy's National Nuclear Security Administration under contract DE-AC04-94AL85000.  The work at Sandia was supported by the U.S. Department of Energy, Office of Science, Materials Sciences and Engineering Division.  Work completed at Purdue was supported by the Department of Energy, Office of Basic Energy Sciences, Division of Materials Sciences and Engineering under Award DE-SC0006671.


\end{document}